\documentclass{ieeeaccess}

\usepackage{cite}
\usepackage{amsmath,amssymb,amsfonts}
\usepackage{algorithmic}
\usepackage{graphicx}
\usepackage{textcomp}
\usepackage{multirow}
\usepackage{caption}
\usepackage{tabularx}
\usepackage{array}
\usepackage{epstopdf}

\begin{document}
\history{Date of publication xxxx 00, 0000, date of current version xxxx 00, 0000.}
\doi{10.1109/ACCESS.2017.DOI}

\title{Measure the Impact of Institution and Paper via Institution-citation Network}
\author{\uppercase{Xiaomei Bai}\authorrefmark{1},
\uppercase{Fuli Zhang}\authorrefmark{2},
\uppercase{Jin Ni}\authorrefmark{3},
\uppercase{Lei Shi}\authorrefmark{4},
\uppercase{Ivan Lee}\authorrefmark{5}}
\address[1]{Computing Center, Anshan Normal University, Anshan 114007, China}
\address[2]{Information Center, Anshan Normal University, Anshan 114007, China}
\address[3]{Adult Education Institute, Anshan Normal University, Anshan 114007, China}
\address[4]{Science and Technology Department, Anshan Normal University, Anshan 114007, China}
\address[5]{School of Information Technology and Mathematical Sciences, University of South Australia, Adelaide SA 5001, Australia}
\tfootnote{This work was partially supported by Liaoning Provincial Key R\&D Guidance Project (2018104021) and Liaoning Provincial Natural Fund Guidance Plan (20180550011).}


\corresp{Corresponding author: Xiaomei Bai (e-mail: xiaomeibai@outlook.com).}

\begin{abstract}
  This paper investigates the impact of institutes and papers over time based on the heterogeneous institution-citation network. A new model, IPRank, is introduced to measure the impact of institution and paper simultaneously. This model utilises the heterogeneous structural measure method to unveil the impact of institution and paper, reflecting the effects of citation, institution, and structural measure. To evaluate the performance, the model first constructs a heterogeneous institution-citation network based on the American Physical Society (APS) dataset. Subsequently, PageRank is used to quantify the impact of institution and paper. Finally, impacts of same institution are merged, and the ranking of institutions and papers is calculated. Experimental results show that the IPRank model better identifies universities that host Nobel Prize laureates, demonstrating that the proposed technique well reflects impactful research.
  \end{abstract}

\begin{keywords}
Institution impact, paper impact, institution-citation network.
\end{keywords}

\titlepgskip=-15pt

\maketitle

\section{Introduction}
\label{sec:introduction}
\begin{figure*}[!htbp]
  \centering
   \includegraphics[width=0.7\linewidth]{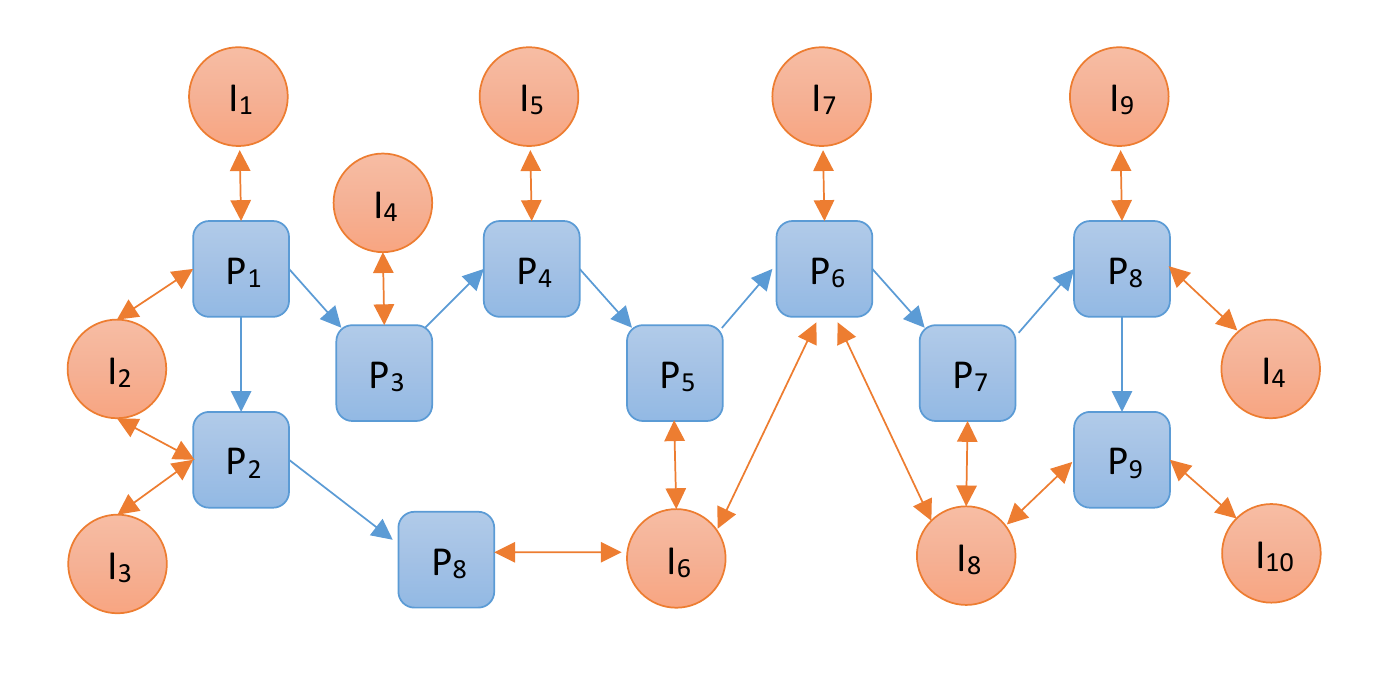}
  \caption{An example of heterogeneous institution-citation network.}
  \label{figure1}
  \end{figure*}

\PARstart{S}{cientific} impact is evaluated at different levels, ranging from high level at national and institutional scales to low level at researcher and paper scales~\cite{bornmann2013universities,bai2016identifying,bai2019predicting}. Many studies focus on scientific impact measure, scholarly network analysis, and success of science~\cite{bao2018identifying,bol2018matthew,lee2015uncovering,iacovacci2016functional,renoust2017multiplex}. While many of these studies explore scientific impact at a particular timeframe, there's a growing interest in understanding the evolution of scientific impact in "science of science"~\cite{rouse2018modeling,fernandez2018questioning}. For scientific impact measurement, citation network is a often used technique~\cite{massucci2019measuring,wang2013ranking}, whereas heterogeneous scholarly network has attracted growing attention recently~\cite{zhou2007co,liang2016scientific}. Quantifying scientific impact in the heterogeneous scholarly network is closely related to structural measure, citation analysis and behavioral complexity. A subset of heterogeneous scholarly networks is the evolving network of institution and paper over time, which forms the structural foundation for advancing scientific discoveries, gauging scientists' performances, ranking universities, and allocating funding. A heterogeneous scholarly network relationship is shown in Figure~\ref{figure1}. $I_{1}$-$I_{10}$ represent research institutes and $P_{1}$-$P_{9}$ represent papers. In Figure~\ref{figure1}, paper $P_{1}$ cites the two papers paper $P_{2}$ and paper $P_{3}$, and the link between two papers points to its reference. The signed institutions of paper $P_{1}$ include institution $I_{1}$ and institution $I_{2}$, the bi-directional links represent the relationship between paper and institution, indicating that the institution publishes the paper and the paper belongs to the institution.
Quantifying paper impact is longstanding point of research~\cite{wang2013quantifying,wang2014future,ke2015defining,stegehuis2015predicting,bai2019predicting}. Previous studies have mainly focused on unstructured measures or structured measures~\cite{battiston2017new}. Unstructured measures rely on citations of scholarly papers or Altmetrics, including downloads, views, shares, and citations~\cite{piwowar2013altmetrics}. Citations attracted by a scholarly paper can sometimes be correlated to its age, which favors older publications. Altmetrics are suitable for quantifying the impact of paper in the early stage of publication. However, both metrics are easily manipulated by scholars who can artificially increase the number of citations. Compared to unstructured metrics, structured metrics more adequately quantify the impact of paper. The most representative structured measures are PageRank and HITS algorithms~\cite{chen2007finding,halu2013multiplex,senanayake2015pagerank,zeitlyn2019perception}. PageRank algorithm is often used in homogeneous network such as citation network and co-author network~\cite{ma2008bringing}. HITS algorithm is used in heterogeneous scholarly network such as paper-author network and paper-journal network~\cite{wang2013ranking}.

Quantifying institution impact has always been the focus of scientific researchers~\cite{molinari2008mathematical,docampo2011using,kapur2016ranking,feng2017computational,banadkouki2018ranking,fernandez2018questioning,rouse2018modeling}.
Currently, quantifying institutional impact is limited to unstructured metrics and homogeneous structured metrics. Several unstructured methods are widely recognized such as Academic Ranking of World University (ARWU), QS World University Ranking (QS), Times Higher Education World University Ranking (THE) and Performance Ranking of Scientific Papers for World Universities (NTU)~\cite{fernandez2018questioning,dobrota2016new}. However, unstructured metrics rely heavily on the number of bibliometric indicators. To develop a structured quantitative method to measure the institutional impact, Massucci et al.~\cite{massucci2019measuring} integrated PageRank into the citation network of institutions. However, despite these significant efforts, the correlation between institutional impact and paper impact in heterogeneous scholarly network remains unclear. Possible reasons include: institution impact evaluation is moving from unstructured to structured; compared to evaluating the institution impact in a homogeneous network, evaluating the institution impact in a heterogeneous network is a more complicated task.

Therefore, we develop a quantitative model, IPRank, to improve the understanding of institution and paper impact in the heterogeneous scholarly network. With the unprecedented expansion of publications and the availability of large-scale datasets on publications, institutions and citations, the analysis of institution and paper network and their quantification in heterogeneous network are now possible. In this paper, we address two main questions. First, we construct a heterogeneous institution-citation network and derive the statistical model of institution-citation network, making it possible to simultaneously quantify the impact of institution and scholarly paper. Second, we develop a structured measurement based on the institution-citation network by utilizing PageRank to quantify the impact of institution and scholarly paper.

The rest of this paper is organized as follows. Section \textrm{II} summarizes recent work on the evaluation of institution and paper impact. Section \textrm{III} introduces the proposed IPRank model framework in detail. The experimental results are shown and discussed in Section \textrm{IV}. Section \textrm{V} draws concluding remarks of the study.

\section{Related Work}
Quantifying the impact of scholarly papers has been extensively investigated. Early studies are mostly based on the number of citations. Garfield proposed using citation counts as the measure of scholarly paper impact~\cite{garfield1955citation}, and he also developed Journal Impact Factor (JIF) as the measure of journal impact~\cite{garfield1972citation}. Although, citation-based approach has certain limitations, such as the impact factor of different disciplines cannot be unified. Citations as a metric to measure the impact of paper have been controversial, especially due to the existence of questionable citations~\cite{catalini2015incidence}.

To resolve this problem, on-going research has been conducted to explore structured metrics to quantify the paper impact~\cite{chen2007finding,molinari2008new,su2011prestigerank,london2015local,liang2016scientific}. These studies are mostly based on scholarly networks, including homogeneous networks (citation network of paper, citation network of institution, and co-author network) and heterogeneous networks (paper-author network, paper-venue network, and author-venue network). Chen et al. \cite{chen2007finding} found scientific gems with Google's PageRank algorithm via citation network. The reason behind it is that important papers attract more citations, including citing paper with high importance, which increase the importance of the cited papers. On the basis of this work, Jiang et al. \cite{jiang2012towards} integrated mutual reinforcement relationships based on the three homogeneous networks and the three heterogeneous networks by applying PageRank and HITS algorithm. Subsequently, Wang et al.~\cite{wang2013ranking} measured the impact of paper by exploiting citations, authors, journals and time information via homogeneous scholarly network and the heterogeneous scholarly network mentioned above. Compared to the work of Jiang et al., Wang et al.~\cite{wang2013ranking} introduced time feature to evaluate the impact of paper, and favored recent scholarly papers to higher scores. Inspired by the work of Wang et al.~\cite{wang2013ranking} and Ioannidis~\cite{ioannidis2015generalized}, Bai et al. \cite{bai2016identifying} proposed COIRank to measure the impact of paper by identifying anomalous citation patterns to adjust citation weights. Liang et al.~\cite{liang2016scientific} proposed a novel mutual ranking algorithm based on the heterogeneous academic hypernetwork by employing the mutual reinforcement relationship. Bai et al. \cite{bai2018quantifying} developed a higher-order weighted quantum PageRank algorithm based on the behaviour of multiple step citation flow. The citation dynamics with higher-order dependencies reveal the actual impact, and better distinguish the impact from self-citation.

Compared to the evaluation of paper impact, quantification of institutional impact is more complicated ~\cite{molinari2008mathematical,perianes2015multiplicative,dobrota2016arwu,perianes2018impact}. Previous metrics are mainly based on statistics of features, including researcher-based features (staff winning Nobel Prizes, number of highly cited researchers, international collaboration), paper-based features (article published in Nature and Science, article index, number of publications, high quality publications, normalized impact, excellence rate, co-publications), institution-based features (university-industry co-publications), and other features such as availability of research funding and graduation rates~\cite{daraio2015rankings,kapur2016ranking,frenken2017drives}. These features are relatively easy to obtain, and they reflect the impact of institution. However, these quantitative indicators have certain drawbacks. Therefore, the structured metrics are investigated to quantify the impact of institution~\cite{bai2017role,massucci2019measuring}. Bai et al. \cite{bai2017role} first explored the conflict of interest (COI) relationships to discover negative citations and weaken the associated citation strength. Furthermore, PageRank and HITS algorithms were utilized to measure the impact of papers based on citation network, paper-author network and paper-journal network. Finally, the institutional impact was calculated by the impact of all publications in this institution. Massucci et al. \cite{massucci2019measuring} studied the citation patterns among university and used the PageRank algorithm based on the citation network between institutions. In their study, the citation relationships between papers are converted into the citation relationships between signed institutions of papers. However, the citation relationships between any two papers is one to one, and since a paper can signed by multiple institutions, the citation relationships between institutions are more complicated.

\section{Methods}
\subsection{Data sources and data pre-processing}
Our experiments are based on the American Physical Society (APS) dataset, which consists of all papers published in Physical Review from 1894 to 2013, spanning across the following journals: Physical Review A, B, C, D, E, I, L, ST and Review of Modern Physics. This dataset includes title of paper, author's name, author's affiliations, date of publication information, and a list of cited papers.

In this study, we consider papers and institutions that meet the following criteria: (1) Paper and institution details are complete and in the right format.
(2) At least one institution is found for a paper.
(3) The first institution associated to each author is retained.
(4) Each institution retains to the first-level unit. For example, we retain Sloane Physics Laboratory, Yale university as Yale university.
(5) Institutions with same name merge. For example, University of California at Berkeley and California University at Berkeley are merged into University of California, Berkeley. It is worth mentioning that before 1952, the University of California at Berkeley was called the University of California. Therefore, in our research, these two names were unified as the University of California, Berkeley.

Through the above pre-processing, a summary of the basic statistics of the APS dataset from 1894 to 2013 is given in Table~\ref{t1}. The entire APS dataset from 1894 to 2013 is used to quantify the long-term impact of institution and paper. Correspondingly, for examining the short-term impact of institution and paper, we summarize the information of the APS dataset during different time periods, also as shown in Table~\ref{t1}. We choose a five-year period to quantify the impact of institution and paper, mainly referring to the Global Ranking of Academic Subjects (ARWU-GRAS) ranking institutions~\cite{massucci2019measuring}. Except for counting the number of papers, the number of institutions, the number of links between papers, and the number of links between papers and institutions, we count the number of references of papers published, including papers published from 1894 to 2013. These references are also used to quantify the short-term impact of institution and paper. The reason is that the literature cited at any time is attributed to the impact of institutions during this period. For instance, to quantify the impact of an institution from 2009 to 2013, we need to construct an institution-citation network, which contain papers published during this time period, references of these papers, and related institutions. A detailed introduction of institution-citation network is covered in the next section.
\begin{table*}[hbt!]
  \centering
  \caption{Statistical summary of the APS dataset for different time periods.}
  \begin{tabular}{lrrrrr}
    \hline
                                                         &1894-2013&1994-1998&1999-2003&2004-2008&2009-2013\\
    \hline
    Papers                                               &  516,162& 62,148  &72,294   &87,049   &94,019\\
    Papers and its references                            &  541,448&151,286  &191,525  &247,416  &295,151\\
    institutions                                         &  227,031&65,046  &92,035  &125,253  &154,023\\
    links between papers                                 &6,040,030&656,203&887,790&1,249,273&1,564,650\\
    links between papers and institutions                &1,057,808&240,220 &344,503 &517,523  &706,147\\
    \hline
  \end{tabular}
  \label{t1}
\end{table*}

\subsection{IPRank model framework}
\begin{figure*}[htbp]
  \centering
  \includegraphics[width=0.8\linewidth]{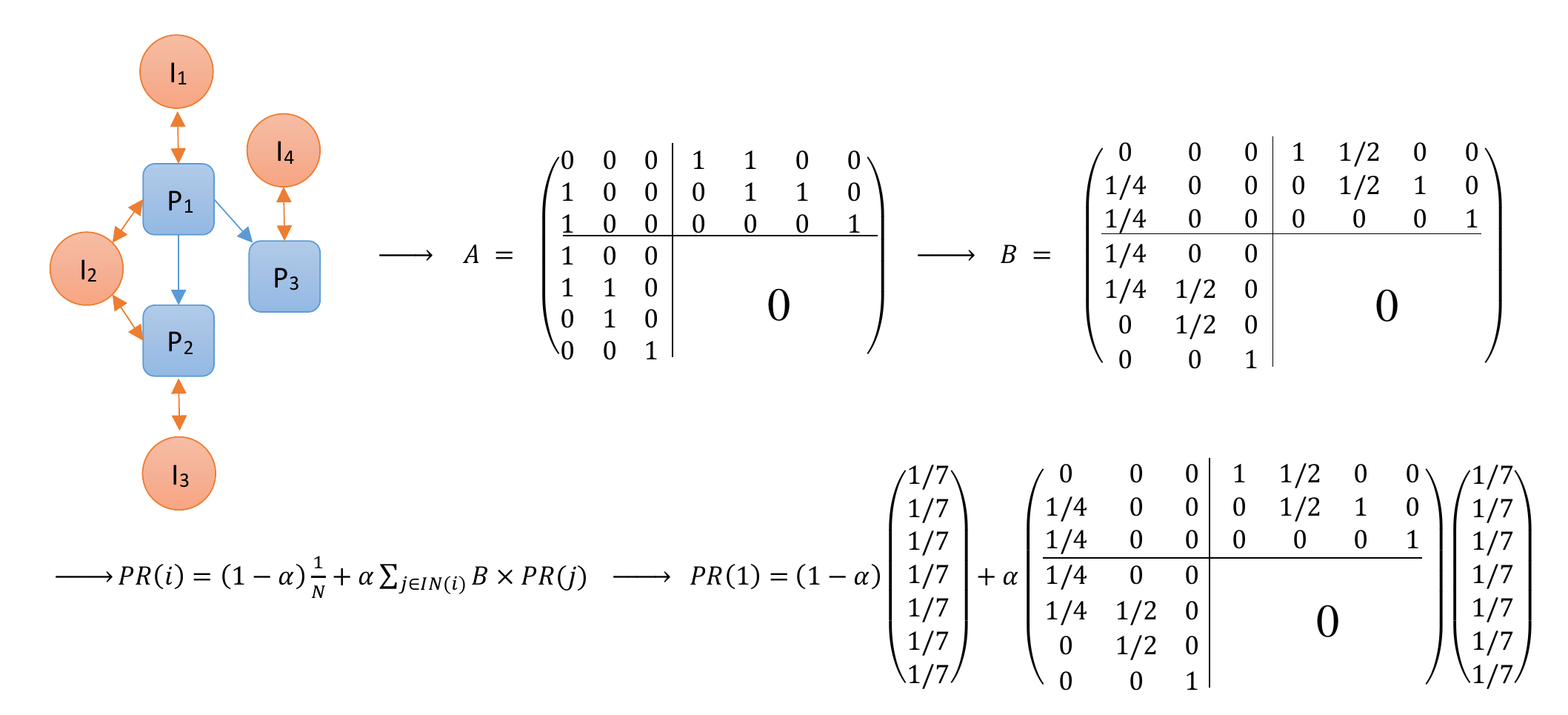}
  \caption{IPRank model.}
  \label{figure2}
\end{figure*}

In this section, we introduce the IPRank model (see Figure \ref{figure2}), which is a PageRank based model for quantifying the impact of institution and scholarly paper. The framework firstly constructs the institution-citation network. PageRank algorithm is then used to quantify the impact of institution and paper. Finally, we merge the impact of institutions, and rank institutions and papers.

\subsubsection{Constructed institution-citation network}
There is a good deal of literature in information science dealing with the citation network between papers~\cite{chen2007finding,bai2016identifying} and the citation network between institutions~\cite{massucci2019measuring} to quantify the impact of paper and the impact of institution. However, to our knowledge, no detailed construction of an actual institution-citation network has been attempted in the past. In this paper, the institution-citation network is a heterogeneous and directed scholarly network, consisting of two categories of nodes: institution and paper. In additions, there are two types of links: one is the citation link between scholarly papers, the other is the link between institution and paper.

Given a set of institutions $I={I_{1}, I_{2},...,I_{m}}$ and a set of scholarly papers $P={P_{1}, P_{2},...,P_{n}}$. Let $E_{PP}$ denote the citations between scholarly papers, $E_{PI}$ denote the relationship between papers and institutions. The heterogeneous institution-citation network can be represented as a graph $G=(I \bigcup P, E_{PP}\bigcup E_{PI})$. For an institution-citation network with $m$ institutions and $n$ papers, graph $G$ can be represented by adjacency matrix $A$:
\begin{equation}\label{eq:e1}
\left(
  \begin{array}{cc}
    A_{PP} & A_{PI}\\
    A_{IP} & 0\\
  \end{array}
\right)
\end{equation}
where $A_{PP}$ represent the citation matrix between papers, $A_{PI}$ and $A_{IP}$ represent the links between institutions and papers. $A_{PI} = A^{T}_{IP}$, since the links between institutions and papers are symmetric.

\subsubsection{IPRank Model}
The motivation of our method is described as follows:
(1) If a scholarly paper is cited by many other publications, it means that the paper has high importance.
(2) If a scholarly paper with a high importance is linked to other papers, the importance of the linked papers will increase accordingly.
(3) If an institution publishes many papers and these papers are cited by many other papers, it means that the institution has high importance.
(4) If a scholarly paper with a high importance is linked to an institution, the importance of the linked institution will increase accordingly.

Figure~\ref{figure2} illustrates IPRank model framework by examine the simple situation: given three papers $P_{1}$, $P_{2}$ and $P_{3}$, paper $P_{1}$ with two institutions, $I_{1}$ and $I_{2}$, paper $P_{2}$ with two institutions, $I_{2}$ and $I_{3}$, paper $P_{3}$ with an institution $I_{4}$. Paper $P_{1}$ cites paper $P_{2}$ and paper $P_{3}$, therefore, a simple citation network can be constructed, which is an unweighted directed graph. According to the relationship between $P_{1}$, $P_{2}$, $P_{3}$ and $I_{1}$, $I_{2}$, $I_{3}$ and $I_{4}$, the links between them can be added to the citation network, thus, a simple institution-citation network (graph $G$) are constructed.

Let $A$ denote the adjacency matrix of $G$, and let $B$ denote the transition probability matrix of $A$. The institution-citation network can be represented by a stochastic matrix $PR$. For a source $i$, the PageRank vector $PR$ is defined as the unique solution of the following formula:
\begin{equation}\label{eq:e2}
  PR(i)=(1-\alpha)\frac{1}{N}+\alpha \sum_{j\in IN(i)}B\times PR(j)
\end{equation}
where $PR(i)$ represents the importance of the node $i$ in the institution-citation network, $\alpha$ (the teleport probability) is a constant between 0 and 1, and is set as 0.85 in our experiments. The value of $\alpha$ parameter refers to the original Google PageRank algorithm~\cite{chen2007finding}. $N$ represents the number of nodes in institution-citation network. $j$ is the adjacent node of $i$, and $j\in IN(i)$ indicates that node $j$ is the indegree of node $i$. $PR(j)$ represents the importance of the node $j$. The linear algebraic definition of PageRank is equivalent to simulating a random walk. Start from the source $i$, with probability (1-$\alpha$), skip to a same chosen neighbor of the current node, or with probability $\alpha$ stop at the current node. According to Equation (\ref{eq:e2}), we finally obtain the prestige scores of institutions and papers in the heterogeneous network.
The pseudocode of IPRank model is listed in ALGORITHM 1.
\begin{table}[hbt!]
  \centering
  \begin{tabular}{l}
    \hline
   ALGORITHM 1: Rank institution and paper\\
    \hline
    Input: Matrix $A_{PP} \in R^{n \times n}$, Matrix $A_{PI} \in R^{n \times m}$,\\
    \quad \quad \quad Matrix $A_{IP} \in R^{m \times n}$\\
    Output: Scores of PR(i)\\
    Initialize Matrix $A$;\\
    Compute transition probability matrix $B$;\\
    Initialize scores of PR(i);\\
    for node $i$ in institution-citation network do\\
         \quad step 1: Calculate scores of PR(i) according to Eq.(2);\\
         \quad step 2: Update scores of PR(i);\\
    end\\
    Iterate step 1 and step 2 until convergence;\\
    Return scores of PR(i);\\
    \hline
  \end{tabular}
  \label{}
\end{table}

The importance of institution and the importance of scholarly paper are their $PR$ values in the institution-citation network. As expected, papers $P_{2}$ and $P_{3}$ are cited by paper $P_{1}$, and paper $P_{3}$ only belongs to institution $I_{4}$, Compared to paper $P_{3}$, paper $P_{2}$ belongs two institutions: $I_{2}$ and $I_{3}$, therefore, $I_{4}$ is the most influential institution among the four institutions. Only paper $P_{1}$ is not cited by other papers in the three papers, therefore, the prestige score of paper $P_{1}$ is the lowest in the three papers. Since institution $I_{1}$ only links paper $P_{1}$, and paper $P_{1}$ with a low prestige score, therefore, the score of the institution $I_{1}$ is the lowest among four institutions. Paper $P_{1}$ and paper $P_{2}$ belong to two institutions, and they share a same institution $I_{2}$. Since paper $P_{1}$ cites paper $P_{2}$, the importance of paper $P_{2}$ is higher than the importance of paper $P_{1}$. Similarly, paper $P_{2}$ and paper $P_{3}$ are also cited by paper $P_{1}$, since paper $P_{2}$ signs two institutions $I_{2}$ and $I_{3}$, and paper $P_{3}$ only signs one institution $I_{4}$, therefore, the importance of institution $I_{4}$ is higher than the importance of institution $I_{3}$.

\section{Results}
We compare the similarity of institution ranking between IPRank and IRank \cite{massucci2019measuring}. Both algorithms can be classified as structured metrics; however, the IPRrank is based on the heterogeneous institution-citation network whereas the IRank is based on the homogeneous citation network between institutions.
Table~\ref{t3} shows the Spearman correlation coefficient between IPRank and IRank.
\begin{table*}[htb]
  \centering
  \caption{Spearman correlation coefficient between IPRank and IRank for top $N$ institutions.}
  \begin{tabular}{lrrrrr}
    \hline
    top $N$&1894-2013&1994-1998&1999-2003&2004-2008&2009-2013\\
    top 10 &0.88     &0.90     &0.75     &0.93     &0.99     \\
    top 20 &0.76     &0.87     &0.86     &0.71     &0.88     \\
    top 30 &0.83     &0.89     &0.62     &0.76     &0.92     \\
    top 40 &0.77     &0.93     &0.75     &0.79     &0.90     \\
    top 50 &0.73     &0.92     &0.83     &0.82     &0.87     \\
    top 60 &0.75     &0.90     &0.85     &0.85     &0.89     \\
    top 70 &0.74     &0.90     &0.87     &0.88     &0.89     \\
    top 80 &0.74     &0.92     &0.89     &0.89     &0.85     \\
    top 90 &0.75     &0.90     &0.91     &0.89     &0.84     \\
    top 100&0.77     &0.90     &0.92     &0.91     &0.85     \\
    \hline
  \end{tabular}
  \label{t3}
\end{table*}

According to Table~\ref{t3}, we observe a high correlation between IPRank and IRank for top 10 - top 100 institutions. In terms of the long-term impact of the institutions, the Spearman correlation coefficient between IPRank and IRank ranges from 0.73 to 0.88 for top 10 - top 100 ranked institutions. Especially, for top 10 institutions, the Spearman correlation coefficient between IPRank and IRank is the highest reaching 0.88. In terms of the short-term impact of the institutions, the Spearman correlation coefficient of the two algorithms changes relatively little, and ranges from 0.87 to 0.93 between 1994 and 1998. Compare to the period from 1994 to 1998, in the two time periods: 1999 to 2003 and 2004 to 2008, the correlation coefficient changed relatively large, from 0.62 to 0.92 and 0.71 to 0.93, respectively. In the five years between 2009 and 2013, the Spearman correlation coefficient between IPRank and IRank is the highest for top 10 institutions reaching 0.99, and the lowest for top 90 institutions reaching 0.84.

\begin{table*}[htb]
  \centering
  \caption{ Spearman correlation coefficient between IPRank and IRank for top $N$ papers.}
  \begin{tabular}{lrrrrr}
    \hline
    top $N$&1894-2013&1994-1998&1999-2003&2004-2008&2009-2013\\
    top 10 &-0.30    &0.70     &0.75     &-0.18    &0.45     \\
    top 20 &0.38     &0.75     &0.68     &0.18     &0.44     \\
    top 30 &0.57     &0.76     &0.77     &0.39     &0.50     \\
    top 40 &0.62     &0.79     &0.73     &0.39     &0.38     \\
    top 50 &0.61     &0.78     &0.77     &0.46     &0.41     \\
    top 60 &0.67     &0.80     &0.49     &0.52     &0.41     \\
    top 70 &0.76     &0.71     &0.35     &0.51     &0.46     \\
    top 80 &0.78     &0.58     &0.41     &0.58     &0.47     \\
    top 90 &0.79     &0.62     &0.41     &0.65     &0.22     \\
    top 100&0.77     &0.67     &0.36     &0.73     &0.06     \\
    \hline
  \end{tabular}
  \label{t4}
\end{table*}

We also compare the similarity of paper ranking between IPrank algorithm and IRank algorithm (see Table~\ref{t4}). In terms of long-term paper impact, the correlation coefficient between the two algorithms is generally on the rise for top 10 - top 100 papers, and ranges from -0.30 to 0.79. During the period from 1994 to 1998,  the correlation coefficient between them is higher than 0.58, and they are all positive related. Between 1999 and 2003, for top 10 - top 50 papers, the correlation coefficient between the IPRank and IRank algorithms is positive related, and they are higher than 0.68. During the same period, for top 60 - top 100 papers, the correlation coefficient between is low, and ranges from 0.35 to 0.49. Between 2004 and 2008, the correlation coefficient between the IPRank and IRank algorithms shows an upward trend, and ranges from -0.18 to 0.73 for top 10 - top 100 papers. Between 2009 and 2013, the correlation coefficient is less than or equal to 0.5. It can be seen that the correlation coefficient at different periods is not regular.

To test whether IPRank model correlates with outstanding impact, we rank 35 Nobel Prize papers from 1930 to 2013 on the basis of IPRank and PageRank. To validate of the IPRank model, we compare the rankings based on IPRank and PageRank. Experimental results indicate 80\% Nobel Prize papers rank higher by IPRank than by PageRank. The top ranked Nobel Prize papers are shown in Table~\ref{t5}, and it indicates that IPRank model has a higher correlation with outstanding impact.

\begin{table}[htbp]
  \centering
  \caption{ Comparing the ranking of IPRank and PageRank algorithms for ten Nobel Prize papers.}
  \begin{tabular}{lrr}
    \hline
    DOI of papers      &IPRank&PageRank\\
    \hline
    PhysRev.108.1175   &2     &4\\
    PhysRevLett.45.494 &11    &40\\
    PhysRev.70.460     &31    &34\\
    PhysRev.73.679     &35    &46\\
    PhysRev.131.2766   &38    &52\\
    PhysRevLett.30.1346&66    &115\\
    PhysRevLett.30.1343&69    &107\\
    PhysRevLett.75.3969&74    &198\\
    PhysRev.76.769     &90    &83\\
    PhysRevB.4.3174    &99    &118\\
    \hline
  \end{tabular}
  \label{t5}
\end{table}

Similarly, we check the rankings of the Nobel Prize institutions between 1930 and 2013, which are derived from Nobel Prize papers. Table~\ref{t6} shows the rankings of ten Noble Prize institutions based on IPRank and IRank algorithms. It should be noted that since 1952, University of California has gradually separated from the University of California, Berkeley as an administrative system, no longer as a university. Therefore, for the institution entry University of California, we also renamed it to the University of California, Berkeley.
According to Table~\ref{t6}, we observe that several institutions have the same ranking order, and several other institutions have slightly different rankings. The reason behind it is that the importance of institution is related to the importance of its published scholarly papers. Simultaneously, the importance of an institution will increase if papers published by the institution are cited by other papers. In general, each institution has a large number of linked papers, and the number of linked papers is different for different institutions. Therefore, the ranking difference based on IPRank and IRank algorithms is small for institution ranking. Compared with institutional rankings, the ranking of a paper depends on its impact of citing papers and institution. Therefore, the rankings of papers ranked by the IPRank and PageRank algorithms are quite different.

\begin{table}[htbp]
  \centering
  \caption{ Comparing the ranking of IPRank and PageRank algorithms for ten Nobel Prize institutions.}
  \begin{tabular}{lrr}
    \hline
    Institution                          &IPRank&PageRank\\
    \hline
    University of California, Berkeley   &1     &1       \\
    Harvard University                   &2     &2       \\
    Princeton University                 &3     &3       \\
    University of Chicago                &4     &6       \\
    Cornell University                   &5     &4       \\
    Stanford University                  &6     &5       \\
    Columbia University                  &7     &13      \\
    University of Illinois               &8     &8       \\
    University of Pennsylvania           &10    &7       \\
    Massachusetts Institute of Technology&19    &35       \\

    \hline
  \end{tabular}
  \label{t6}
\end{table}

\begin{figure}[htbp]
  \centering
  \includegraphics[width=1\linewidth]{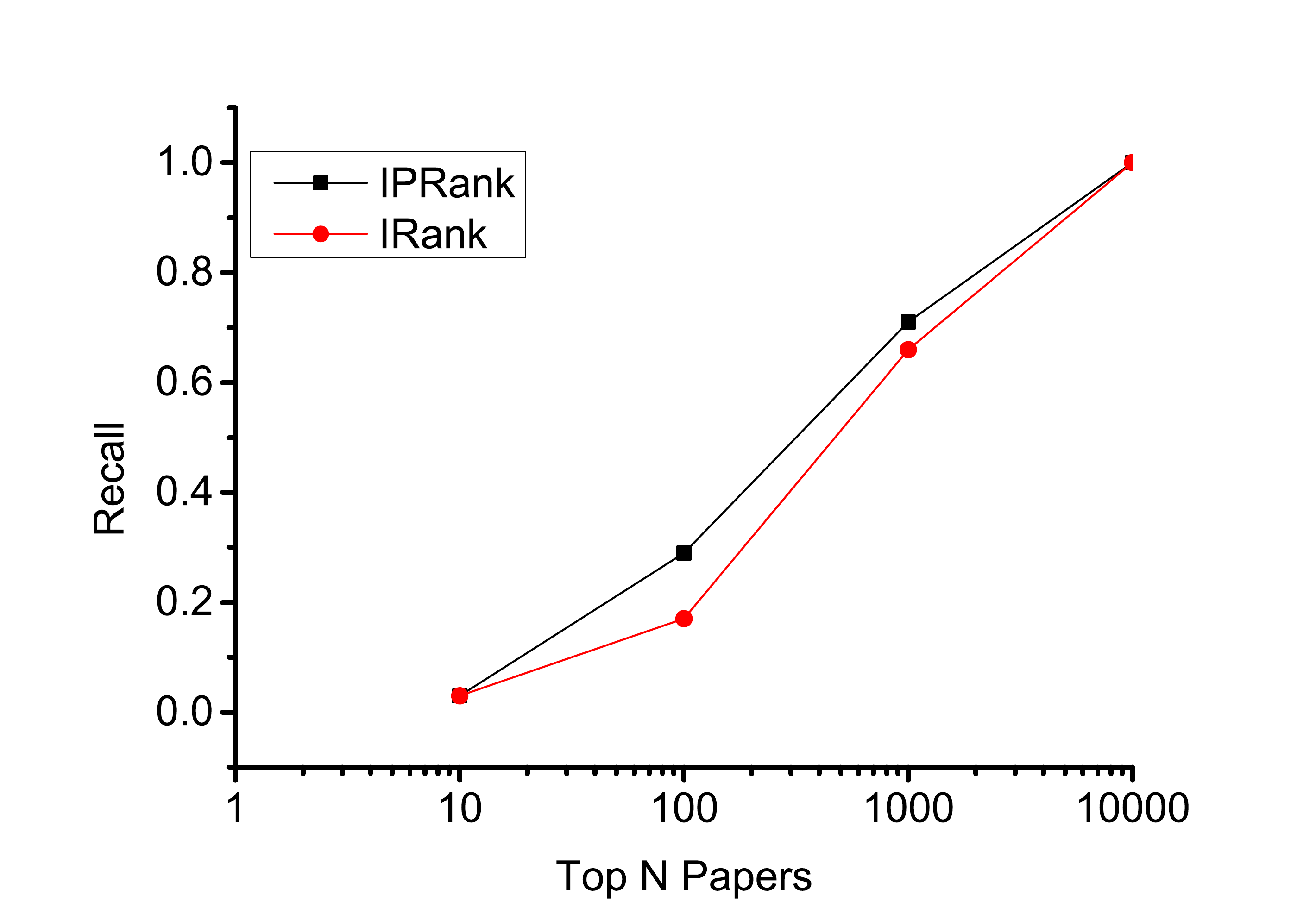}
  \caption{Recall performance for retrieving Nobel Prize papers among top $N$ papers.}
  \label{figure3}
\end{figure}

Figure~\ref{figure3} compares IPRank and PageRank in terms of the recall rates of retrieving 35 Nobel Prize papers among top $N$ papers. It is observed that the IPRank algorithm consistently yields higher recall rates than the PageRank algorithm. Thus, the IPRank algorithm better reflects the impact of Nobel Prize papers.

\begin{figure}[htbp]
  \centering
  \includegraphics[width=0.95\linewidth]{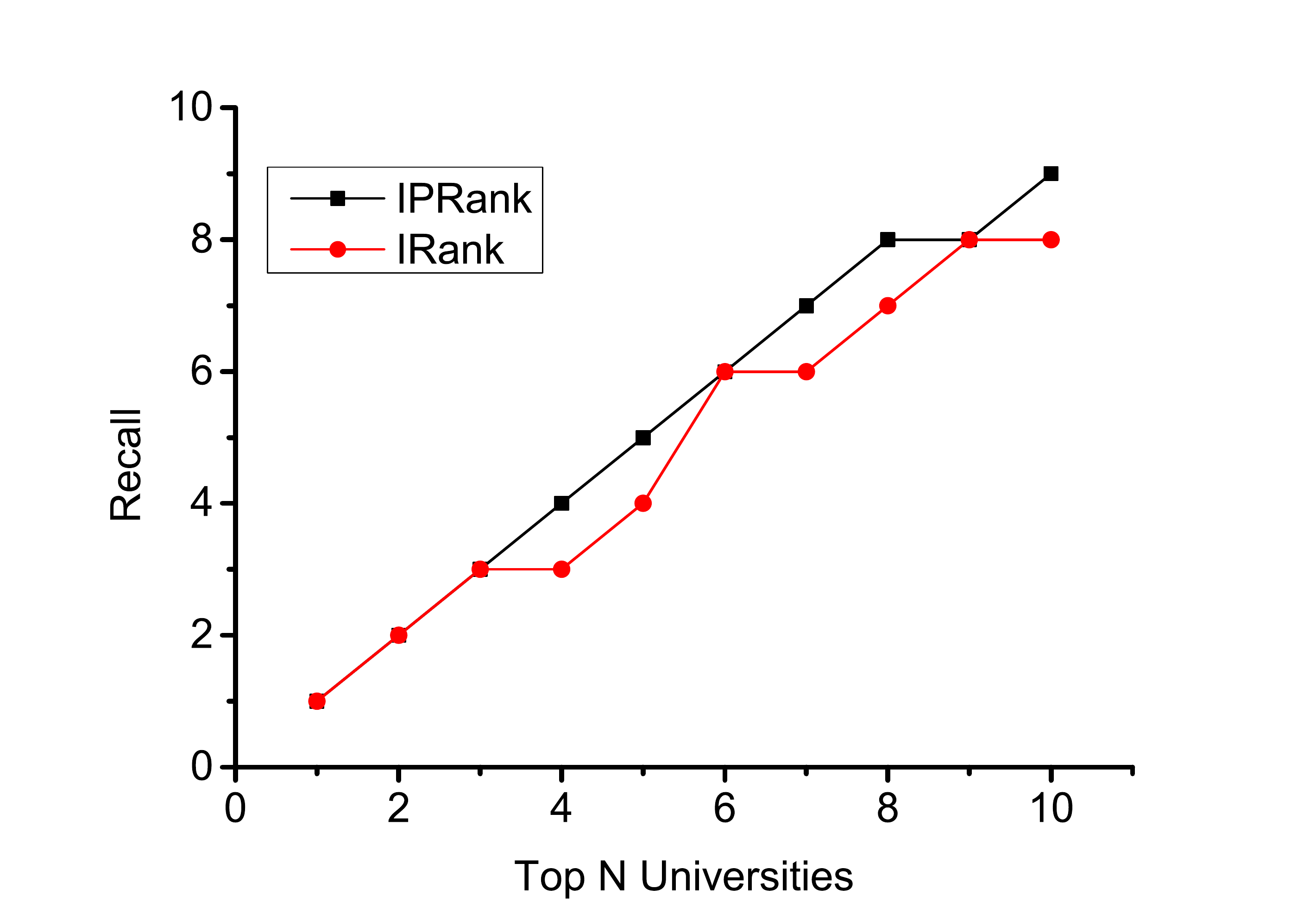}
  \caption{Recall performance for retrieving Nobel Prize universities among top $N$ universities.}
  \label{figure4}
\end{figure}

Figure~\ref{figure4} compares IPRank and IRank in terms of the recall rates of identifying Nobel Prize universities and among top $N$ universities. For top 1 to top 3, top 6 and top 9 universities, both IPRank and IRank contain the same number of Nobel Prize universities. For top 4, top 5, top 7, top 8 and top 10 universities, the IPRank consistently yields higher recall rates than that of IRank, indicating that the IPRank algorithm better reflects the impact of Nobel Prize institutions.

\begin{figure}[htbp]
  \centering
  \includegraphics[width=0.95\linewidth]{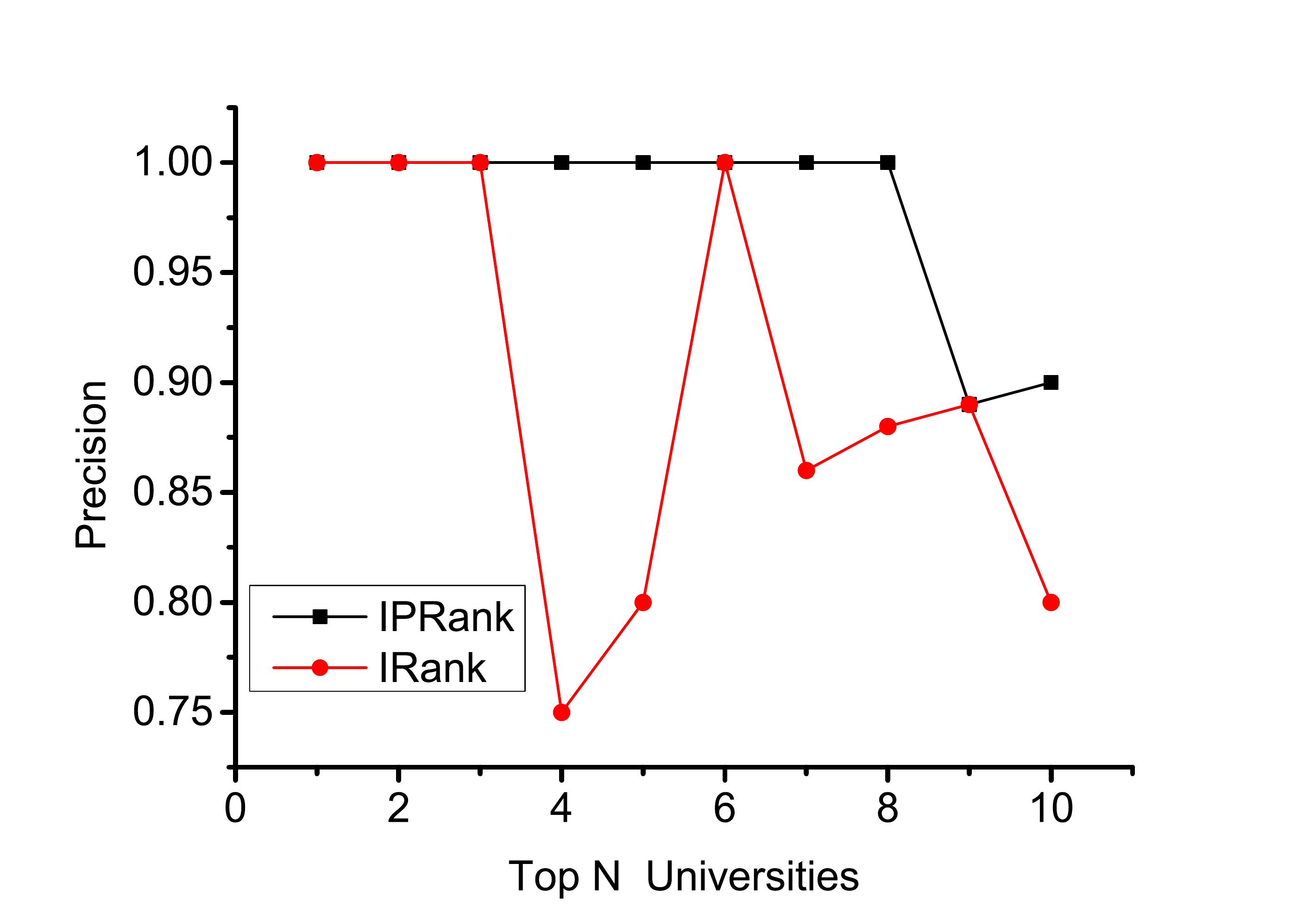}
  \caption{Precision performance for retrieving Nobel Prize universities among top $N$ universities.}
  \label{figure5}
\end{figure}

Figure~\ref{figure5} compares IPRank and IRank in terms of the precision rates of retrieving Nobel Prize universities and among top $N$ universities. From top 1 to top 8 universities, the probability of the number of Nobel Prize universities of IPRank is 1. For top 9 and top 10 universities, the probability of the number of Nobel Prize universities of IPRank is less than 1, and they are 0.88 and 0.90 respectively. In contrast, the probability of the number of Nobel Prize universities of IRank fluctuates greatly and ranges from 0.80 to 0.89. The probability of the number of Nobel Prize universities of the IPRank algorithm is found greater than or equal to the probability using the IRank algorithm.

\section{Conclusion}
This paper investigated a data-driven method to quantify the impact of institution and paper from heterogeneous institution-citation network. Unlike most prior studies that utilised citation network to measure the impact of institution or paper, this paper proposed IPRank to simultaneously quantify the impact of institution and paper in a heterogeneous scholarly network. Experimental results showed that the IPRank model was more representative of the outstanding impact of institution and paper. Compared to the ranking of IPRank and PageRank algorithms for Nobel Prize papers and institutions, IPRank model produced a higher ranking in most cases for identifying Nobel Prize-winning papers and institutions, making it an adequate tool for institutional impact assessment.

\bibliographystyle{IEEEtran}

\EOD

\end{document}